\documentclass[doublecol,figures]{epl2}

\usepackage{grffile}
\usepackage{epstopdf}
\usepackage{amsmath,amssymb,psfrag,color}
\newcommand{\D}{\mathrm{d}}
\newcommand{\E}{\mathrm{e}}
\newcommand{\en}{\mathcal{N}}
\newcommand{\fext}{f_{\mathrm{ext}}}
\newcommand{\fmu}{f_\mu}
\newcommand{\mtau}{\langle \tau \rangle}
\newcommand{\mtp}{\langle \tau_+ \rangle}
\newcommand{\mtm}{\langle \tau_- \rangle}
\newcommand{\mtpm}{\langle \tau_\pm \rangle}
\newcommand{\win}{w_{\mathrm{in}}}
\newcommand{\wout}{w_{\mathrm{out}}}
\newcommand{\Pin}{P_{\mathrm{in}}}
\newcommand{\Pout}{P_{\mathrm{out}}}
\newcommand{\xout}{x_{\mathrm{out}}}
\newcommand{\xin}{x_{\mathrm{in}}}
\newcommand{\Pss}{P_{\mathrm{ss}}}
\newcommand{\Jss}{J_{\mathrm{ss}}}
\newcommand{\Jyss}{J^y_{\mathrm{ss}}}
\newcommand{\Jxss}{J^x_{\mathrm{ss}}}
\newcommand{\vst}{v_{\mathrm{ss}}}
\newcommand{\tps}{\tilde{\tau}_{+}}
\newcommand{\tms}{\tilde{\tau}_{-}}
\newcommand{\vx}{v_{\mathrm{ss}}^x}
\newcommand{\vy}{v_{\mathrm{ss}}^y}
\newcommand{\Ubar}{\overline{U}}

\title{Efficiency of molecular machines with continuous phase space}
\shorttitle{Efficiency of molecular machines\dots}
\author{N. Golubeva\inst{1} \and A. Imparato\inst{1} \and L. Peliti\inst{2}}
\institute{%
  \inst{1} Department of Physics and Astronomy, University of Aarhus, Ny Munkegade, Building 1520, DK--8000 Aarhus C, Denmark\\
  \inst{2} Dipartimento di Scienze Fisiche and Sezione INFN,
Universit\`a ``Federico II'', Complesso Monte S. Angelo, I--80126 Napoli, Italy
}

\abstract{
We consider a molecular machine described as a Brownian particle diffusing in a tilted periodic potential. We evaluate the absorbed and released power of the machine as a function of the applied molecular and chemical forces, by using the fact that the times for completing a cycle in the forward and the backward direction have the same distribution, and that the ratio of the corresponding splitting probabilities can be simply expressed as a function of the applied force. We explicitly evaluate the efficiency at maximum power for a simple sawtooth potential. We also obtain the efficiency at maximum power for a broad class of 2-D models of a Brownian machine and find that loosely coupled machines operate with a smaller efficiency at maximum power than their strongly coupled counterparts.}

\begin{document}

\maketitle

\section{Introduction} Understanding the efficiency of the free-energy transduction in molecular motors and more generally nano-machines requires a different set of concepts than those used in the theory of Carnot engines in macroscopic thermodynamics~\cite{Hill,PhysRevLett.72.2656,esposito09,esposito10}. They work in an environment at a constant temperature, and therefore their maximum efficiency is equal to 1, and is reached when their output power vanishes. It is more interesting, therefore, to understand the behavior of their efficiency as a function of their output power, and in particular their efficiency at maximum power (EMP). The issue of the EMP in nano-machines has recently attracted considerable interest, see, e.g. \cite{Esposito2012,Gaveau2010} and references therein. In particular the EMP in molecular motors has been investigated in \cite{Schmiedl2008a,SeifertPRL}.

Since these systems are subject to fluctuating interactions with their environment, they must be modeled as stochastic processes. Their description can be performed at different levels of sophistication: as Markov chains with discrete states, or as diffusion processes with continuous states (or as a combination in which some degrees of freedom are continuous and other discrete). In all cases, it is important to take into account the constraints that microscopic reversibility and thermodynamic consistency impose on the dynamics. The de~Donder relation~\cite{DeDonder}, which connects the ratio of the reaction rates in the forward and backward direction with the free-energy difference, expresses these constraints for a Markov chain model of a nano-machine. It has been recently shown~\cite{SeifertPRL} that this relation allows to derive a rather elegant expression for the output power of the system as a function of the free-energy imbalance which keeps it moving and of a single time scale which depends on its detailed dynamics. This has allowed for an investigation of the EMP for a set of discrete models of nano-machines.

One can describe a molecular machine with a continuous phase space as a Brownian particle diffusing in a potential~\cite{PhysRevLett.72.2656}. The microscopic constraints are expressed in this case by the Einstein relations between the kinetic coefficients and the noise correlations. A simple description is obtained by considering a tilted periodic potential. In this case one coordinate ($x$) describes the spatial location of the motor, and the second coordinate ($y$) describes the advancement of a chemical reaction, such as ATP hydrolysis. Then the tilt in the $x$ direction describes the external mechanical force applied on the system, while the tilt in the $y$ direction describes the thermodynamic imbalance between the ``fuel'' (ATP) and the ``exhausts'' (ADP + P$_{\mathrm{i}}$). In this work we wish to evaluate the output power of such a system, and its corresponding efficiency.

We will introduce and discuss a general formalism that allows us to obtain the absorbed and released power of model motors as a function of the applied mechanical and chemical forces, for any choice of the underlying potential in the 1-D case. We proceed by evaluating explicitly the efficiency at maximum power for a sawtooth potential in 1-D, as well as for a specific class of 2-D potentials.

\section{Unicyclic machines} We shall first discuss unicyclic machines, i.e.,  machines in which the mechanical and the chemical cycles are tightly bound, so that the system is constrained to move along a one-dimensional trajectory in the $(x,y)$ space. This situation holds when the potential driving the particle is much smaller along a trajectory $y(x)$ than far from it. In this case we can neglect the fluctuations, say, of $y$ at fixed $x$ and describe the system as a Brownian particle moving in a tilted one-dimensional potential $U(x)=U_0(x)-f x$, where $U_0(x)$ is a periodic function with period $L$, and $f$ is a generalized external force coupled with the position $x$ of the machine along the cycle.
The force $f$ can be considered  as the sum of two contributions, $f=-\fext+\fmu$ where $\fext>0$ is the external  force  opposing the motion and $\fmu>0$ is the chemical driving force. Despite its simplicity, this model is general enough to describe a large class of model motors: as an example, the two-level ratchet motors can be described as a Brownian particle diffusing  in an effective one-dimensional tilted potential \cite{PhysRevLett.72.2652,RevModPhys.69.1269}.

When the machine completes a cycle in the positive direction, it dissipates the energy $\win=\fmu L$ from the input reservoir and delivers the output work $\wout=\fext L$ to the environment.  The net energy dissipated after completing a cycle is given by $f L=\win-\wout$. The mean times to complete a full cycle in the positive and negative directions are denoted by $\mtp$ and $\mtm$, respectively. The probability distribution function (PDF) $P(x,t)$ that the particle is at position $x$ at time $t$ satisfies the Fokker-Planck equation
\begin{equation}
  \label{eq:FP}
  \frac{\partial P}{\partial t}+\frac{\partial J}{\partial x}=0,
\end{equation}
where the  probability current is given by $J(x,t)=-\Gamma \left[ \partial_{x}U\,P + T \partial_{x}P \right]$. Here $\Gamma$ is the mobility, and we set $k_\mathrm{B}=1$ throughout.

With a straightforward calculation, one finds that the steady-state probability distribution, satisfying the boundary conditions $\Pss(0)=\Pss(L)$, is given by \cite{vanKampen}
\begin{equation}
\Pss(x)=\en \frac{\E^{-\beta U(x)}}{T}\left[\frac{I(L)}{1-\exp\left(-\beta f L\right)}-I(x)\right],
\label{pss}
\end{equation} 
where $I(x)=\int_0^x \D z \exp\left[\beta U(z)\right]$, and $\en$ is a normalization constant. Thus, one can obtain the steady state current $\Jss=\Gamma\en$, and the steady state velocity $\vst=\Jss L =\Gamma L\en$.

\section{Mean times and splitting probabilities} We shall now consider the expression for the splitting probabilities, i.e., the  probabilities $p_+$ and $p_-$ that the particle completes a cycle in the positive or negative direction, given that it starts at $x=0$ at the initial time. Given the fact that the mean times $\mtpm$ are equal, we can write down a simple expression for the steady-state velocity. 

The PDF $P_0(x,t)$ that a particle starting from $x=0$ at time $t=0$ is found at position $x$ at time $t$, without having completed a cycle, satisfies eq.~(\ref{eq:FP}) with the absorbing boundary conditions
\begin{equation}
  \label{eq:BC}
  P_0(-L,t)=P_0(L,t)=0,
\end{equation}
and the initial condition $P_0(x,0)=\delta(x)$. We want to characterize the PDFs $P_\pm(t)$  of the escape times for cycles in the positive or in the negative direction,  which  read
\begin{equation}
  \label{eq:P_pm}
  P_\pm(t)=\frac{J_\pm(t)}{\int_0^\infty J_\pm(t' ) \text{d}t' },
\end{equation}
where $J_\pm$ are the (positive) probability currents at the boundaries
\begin{equation}
  \label{eq:J_pm}
  J_\pm(t)=\mp \Gamma T \left. \frac{\partial P_0(x,t)}{\partial x} \right|_{x=\pm L},
\end{equation}
where we have exploited the boundary conditions (\ref{eq:BC}). It can be shown by directly solving the Fokker-Planck equation that the currents obey the relation \cite{ISI:000202876400012}
\begin{equation}
  \label{eq:J_p/J_m}
  \frac{J_+(t)}{J_-(t)}=\E^{fL/T}.
\end{equation}
This expression enables us to evaluate the thermodynamic quantities of systems with continuous phase space in terms of a single microscopic timescale as described below. Combining equations~(\ref{eq:P_pm}) and (\ref{eq:J_p/J_m}), we find that $P_\pm(t)$ are equal, and hence
\begin{equation}
  \label{eq:mtpm}
  \mtp=\mtm =\int_0^\infty  t' P_\pm(t') \, \D t'\equiv \mtau.
\end{equation}
We have thus the somehow surprising result that the characteristic times for performing a cycle ``upstream'' are the same as for a ``downstream'' cycle. The splitting probabilities $p_\pm$ are given by
\begin{equation}
  \label{eq:p_pm}
  p_\pm=\frac{\int_0^\infty J_\pm(t) \text{dt}}{\int_0^\infty (J_+(t)+J_-(t)) \text{dt}}=\frac{1}{1+\E^{\mp fL/T}}.
\end{equation}
and satisfy the relation ${p_+}/{p_-}=\E^{fL/T}$. As pointed out by van~Kampen~\cite[p.317]{vanKampen}, this relation can also be directly derived, without explicitly solving the Fokker-Planck equation, by following, e.g., the method reported in~\cite[p.142f]{Gardiner}.

We are now able to establish a connection between the steady state process and the typical escape time $\mtau$, as obtained by solving the Fokker-Planck equation with absorbing boundaries. One obtains indeed the following relation
\begin{equation}
\vst=\frac{L}{\mtau}(p_+-p_-)=\frac{L}{\mtau}\frac{1-\E^{- fL/T}}{1+\E^{- fL/T}},
\label{eq:vst}
\end{equation}  
which yields the value of $\mtau$ once the steady-state PDF (and thus $\vst$) has been obtained from eq.~(\ref{pss}). This relation is verified by numerical simulations in the following.

According to (\ref{eq:mtpm}), the typical times for forward and backward motion are equal, and hence the power associated with motion in the positive and negative direction is determined by the splitting probabilities. The input power in the steady state is given by
\begin{equation}
  \label{eq:P_in}
  \Pin \equiv \fmu \vst = \frac{\win}{\mtau}(p_+-p_-)=\frac{\fmu L}{\mtau}\frac{1-\E^{- fL/T}}{1+\E^{- fL/T}}.
\end{equation}
Similarly, the power delivered by the motor is given by
\begin{equation}
  \label{eq:P_out}
  \Pout \equiv \fext \vst = \frac{\wout}{\mtau}(p_+-p_-)=\frac{\fext L}{\mtau}\frac{1-\E^{- fL/T}}{1+\E^{- fL/T}}.
\end{equation}
Equations (\ref{eq:P_in}--\ref{eq:P_out}) represent the continuous phase space analog of equations (3--4) in \cite{SeifertPRL}. The thermodynamic efficiency is defined as $\eta\equiv \Pout/\Pin = \wout/\win=\fext/\fmu$ and is bounded by $0<\eta<1$. At thermodynamic equilibrium, $f=0$, the efficiency attains its maximum value $\eta=1$, but the output power vanishes, which is a well-known result for Carnot machines.

Let us define $\tps=\mtau/p_+$ and $\tms=\mtau/p_-$, which correspond to $\tau^{\pm}$ in~\cite{SeifertPRL}. There, assuming an Arrhenius expression for these characteristic times, the analysis is carried out under the assumption that the quantities $\xin=-\D\,{\ln(\tps)}/\D\win$ and $\xout=-\D\,{\ln(\tps)}/\D\wout$ are independent of $f$. However, by inverting eq.~(\ref{eq:vst}), one can express $\mtau$ in terms of the other quantities, and thus obtain explicit expressions for $x_{\mathrm{in,out}}$, which are not constant in general, as one can check on an explicit example.

\section{An example: The sawtooth potential}
A simple example to illustrate the above concepts is provided by the sawtooth potential
\begin{equation}
  \label{eq:U_0}
  U_0(x)=\begin{cases} \gamma_0 x, & \text{if}\quad 0<x<a, \\ \delta_1+\gamma_1 x, & \text{if} \quad a<x<L, \end{cases} 
\end{equation}
with $\delta_1=(\gamma_0-\gamma_1)a$, $\gamma_0=\Ubar/a$, $\gamma_1=\Ubar/(a-L)$, and where $\Ubar=U_0(a)$ is the potential maximum. For such a potential, analytical expressions for all the relevant quantities ($\Pss$, and thus $\vst$ and $\mtau$) can be readily obtained.
\begin{figure}[ht]
\center
\psfrag{vm}[ct][ct][1.]{$\vst$}
\psfrag{f}[ct][ct][1.]{$f$}
\includegraphics[width=0.45\textwidth]{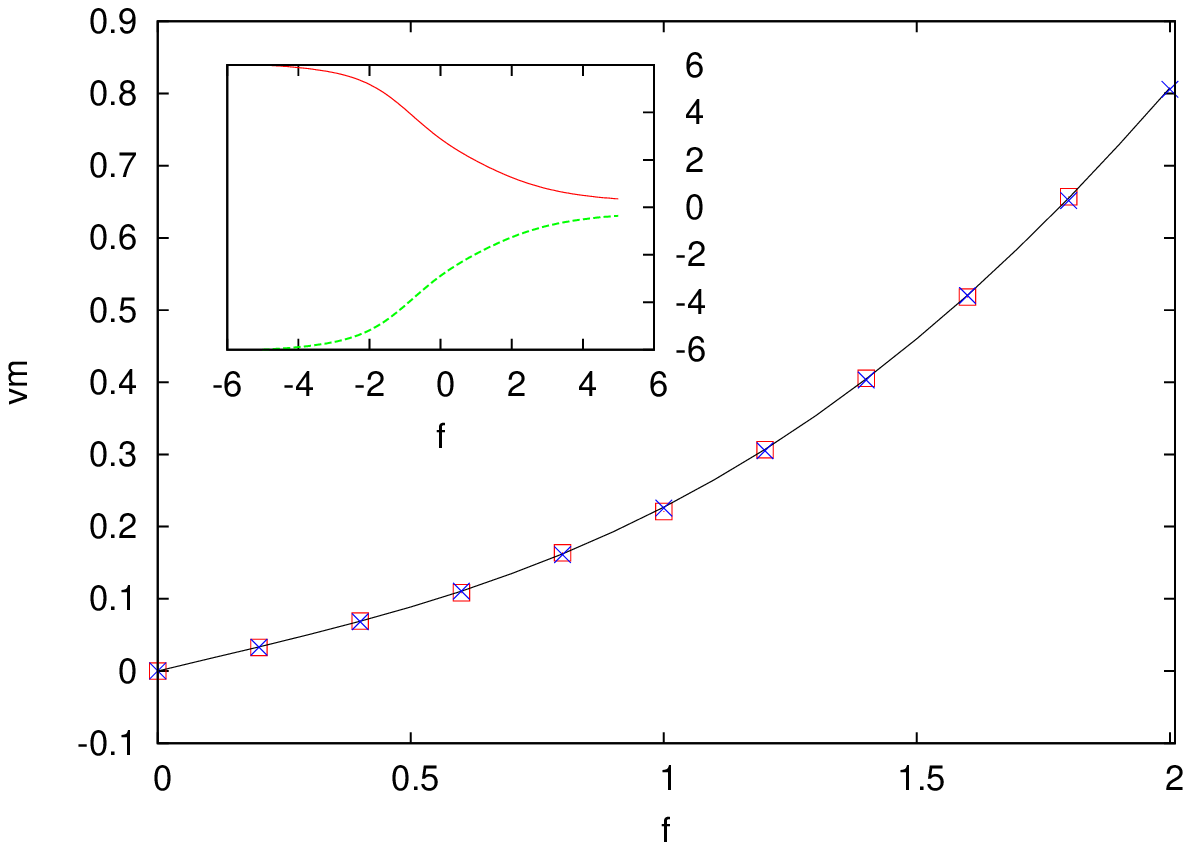}
\caption{(Color online) Steady-state velocity $\vst$ in the potential (\ref{eq:U_0}) as a function of the total force $f$. Comparison between the exact value $\vst=\Jss L$ (full line), where $\Pss$ is given by eq.~(\ref{pss}), and the expressions~(\ref{eq:vst}) (squares), where $\mtau$ is obtained from simulating the escape process with absorbing boundary conditions (\ref{eq:BC}). Crosses: results of numerical simulations for the steady state diffusion with periodic boundary conditions. Inset: Plot of  $\xout$ (dashed line) and $\xin$ (full line) vs.\  the total force $f$. We take  $a=0.35$, $T=0.2$, $\Gamma=L=\Ubar=1$.}
\label{fig_v}
\end{figure}
We first check eq.~(\ref{eq:vst}), by evaluating $\vst$ from the exact relation $\vst=\Jss L$ for different values of $f$. We then simulate the escape process in the potential (\ref{eq:U_0}), with absorbing boundary conditions (\ref{eq:BC}), to obtain an estimate of $\mtau$. Thus, in fig.~\ref{fig_v}, we plot the analytic prediction for $\vst$ and the value obtained by the rhs of eq.~(\ref{eq:vst}), finding an excellent agreement. As a further check, we report in the same figure the results of simulations for the steady state diffusion in the potential (\ref{eq:U_0}) with periodic boundary conditions. In the inset of fig.~\ref{fig_v}, $\xin$ and~$\xout$, as defined above, are plotted vs.\ $f$ for the potential (\ref{eq:U_0}). Such a plot clearly shows that $\xout$ and $\xin$ vary with $f$ and thus with $\wout$ and $\win$, respectively.

\section{Efficiency at Maximum Power (EMP)} 
\begin{figure}[ht]
  \center
\psfrag{win}[ct][ct][1.]{$\win$}
\psfrag{eta}[ct][ct][1.]{$\eta^*$}
\psfrag{legend1}[cr][cr][0.8]{$\lambda_a=0.1$}
\psfrag{legend2}[cr][cr][0.8]{$\lambda_a=0.3$}
\psfrag{legend3}[cr][cr][0.8]{$\lambda_a=0.7$}
  \includegraphics[width=8cm]{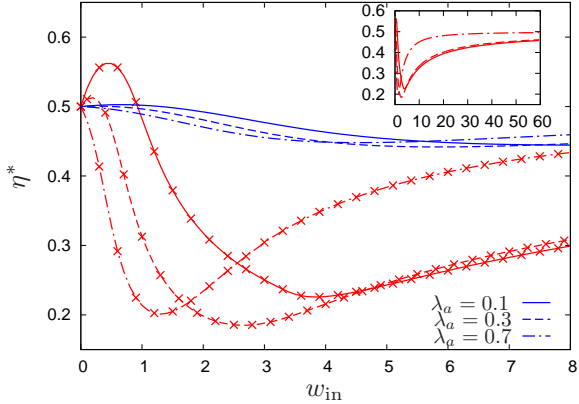}
  \caption{(Color online) EMP $\eta^*$ for maximization with respect to $\wout$ as a function of $\win$ in a tilted sawtooth potential for different values of the temperatures $T$ and the asymmetry parameter $\lambda_a$. The rest of the parameters are as in fig. \ref{fig_v}. Lines: $T=0.4$, Lines and symbols: $T=0.1$. Inset: Plot of $\eta^*$ in the large $\win$ regime, for $T=0.1$. }
  \label{fig:EMP_sawtooth}
\end{figure}

We can now evaluate the EMP for the model machine,  maximizing the output power with respect to $\win$,  $\wout$, or both.  For fixed $\wout$, the maximization condition reads $\partial \Pout/\partial \win=\fext \partial v/\partial \win=0$ giving the optimal input $\win^*=\infty$. Therefore the EMP vanishes for all values of $\wout$. On the other hand, if $\win$ is fixed, the condition $\partial \Pout/\partial \wout=0$ yields the value of the output $\wout^*$ that maximizes the power, $\Pout(\wout^*)=\Pout^*$. The corresponding velocities at maximum power and EMP are denoted by $\vst^*$ and $\eta^*=\wout^*/\win$, respectively. Figure~\ref{fig:EMP_sawtooth} shows the results obtained for the model in the potential (\ref{eq:U_0}). 

It is interesting to remark that $\eta^{*}$ rises above its linear response value $1/2$ for small values of $\win$ and for values of the asymmetry parameter $\lambda_a=a/L<0.5$, and decreases afterwards. In the low temperature regime $(\Ubar-fL)/T\gg 1$ this can be understood by applying the Kramers approximation, since in this limit the system can be described by a one-dimensional Markov process with one single state corresponding to the potential minima. Thus, in this limit our description is approximately equivalent to \cite{SeifertPRL}, and hence $\xout \simeq -\lambda_a$. Expanding in the chemical driving force we obtain
\begin{equation}
  \label{eq:seifert_eta_small_fmu}
  \eta^*=1/2+1/8(1/2+\xout)\win/T+O(f_\mu^2),
\end{equation}
which then explains the behaviour of the EMP for different values of $\lambda_a$ seen in fig.~\ref{fig:EMP_sawtooth}.

Far from equilibrium, $f_\mu \gg \fext$, the Kramers description breaks down, and since  $\vst \simeq \Gamma f$ in this limit, the EMP approaches $1/2$ for $f L \gg \Ubar$ or, equivalently, $\win\to \infty$, see fig.~\ref{fig:EMP_sawtooth}. This behaviour is very different from the behaviour of the discrete system, where as $\win\to \infty$ we have $\eta_{\text{out}}^*\to 1$  if $\xout>0$, or $\eta_{\text{out}}^*\to 0$  if $\xout<0$ \cite{SeifertPRL}. Again, this is a consequence of the fact that $\xout$ is not constant, or in other words that neither the microscopic rate constant in the Kramers rate expression nor the position of the potential minima is independent of $f$.
 
\section{2-D system} We shall now describe the motor as a Brownian particle in a 2-D potential, with a spatial ($x$) and a ``chemical'' ($y$) coordinate. We take the mobilities for the two degrees of freedom to be equal, i.e. $\Gamma_x=\Gamma_y \equiv \Gamma$. We choose an unperturbed potential of the form
\begin{equation}
U_0(x,y)=U_1(z_1(x,y))+U_2(z_2(x,y)),
\label{u2d}
\end{equation} 
where $U_1$ and $U_2$ are periodic potentials with the directions defined by $z_1=n_1x-m_1y$ and $z_2=n_2x-m_2y$ and periods $L_1$, $L_2$, respectively. The ratio $U_1/U_2$ thus represents the coupling strength between the two coordinates. The tilts along the $x$ and $y$ direction are given by $f_x$ and $f_y$, respectively. With this choice for the potential we have $\Pin=f_y \vy$, $\Pout=-f_x \vx$,  $\win=L_y f_y$,  $\wout=L_x f_x$, where $L_x$, $L_y$ are the periods of the potential along the $x$- and $y$-direction, respectively. 

When $U_2=0$, the potential (\ref{u2d}) is the particular case of a washboard potential considered in ref.~\cite{PhysRevLett.72.2656}. By making the ansatz $\Pss(x,y)=\mu p(nx-my)=\mu p(z)$ (and dropping the subscripts for now), where $\mu$ is a normalization constant, the problem becomes one-dimensional. The solution for $p(z)$ is then given by \eqref{pss} with the effective potential $V(z)=U_0(z)-\xi z$, where the effective force becomes $\xi=(f_xn-f_ym)/(n^2+m^2)$. In terms of the stationary current $J=\Gamma \en$ obtained from $p(z)$ the probability currents in the original coordinates become    
\begin{align}
  \Jxss(x,y)&=n \mu J+\Gamma(\alpha f_x + \beta f_y) \Pss(x,y) \\
  \Jxss(x,y)&=- m \mu J+\Gamma(\beta f_x + \gamma f_y)\Pss(x,y),
\end{align}
with $\alpha=m^2/q^2$, $\beta=nm/q^2$, $\gamma=n^2/q^2$, $q^2=n^2+m^2$. At low temperatures we have $J\to 0$ as expected, and the velocities are thus given by
\begin{align}
  \vx=L_x \oint \Jxss \D y = \Gamma( \alpha f_x + \beta f_y) \\
  \vy=L_y \oint \Jyss \D x = \Gamma( \beta f_x + \gamma f_y),
\end{align}
where we have exploited the normalization condition for $\Pss(x,y)$. Hence, in this case the velocities are always linear in the forces at low temperature for \emph{all} values of $f_x$ and $f_y$. The optimizing force becomes $f_x^*=-n/2m f_y$, and hence $\eta^*=1/2$. At higher temperatures, the particle can diffuse out of the pathways introduced by the washboard potential. Hence, the tight coupling is lost, i.e., the steady state velocities $\vx$ and $\vy$ are no longer proportional, and we expect that $\eta^*<1/2$. Thus, $1/2$ is the maximum EMP that can be achieved for a single washboard potential.

When $U_2\neq 0$ the motion of the system can be described along the two independent coordinates $z_1$ and $z_2$. The Fokker-Planck equation decouples and can hence be solved analytically only when $z_1$ and $z_2$ define orthogonal directions. Then, one can evaluate the two independent steady-state PDFs along $z_1$ and $z_2$, as given by eq.~(\ref{pss}), solve the problem and revert to the original coordinates. 
In the following we consider two sawtooth potentials 
 with barrier height $\Ubar_i$ and asymmetry parameters $\lambda_i$. Furthermore, we take $z_1=x-y$, $z_2=x+y$. Since in the following we want to use a Kramers formalism to describe the 2-D diffusion in the low temperature regime, and the sawtooth potentials are not differentiable at their extrema, we approximate such potentials with their Fourier series up to the third order. The exact results for the steady state velocities obtained with the approximated potentials are however very similar to the ones obtained with  the original sawtooth potentials.

\begin{figure}[h]
\psfrag{U1/U2}[ct][ct][1.]{$\delta$}
\psfrag{fy}[ct][ct][1.]{$f_y$}
  \centering
  \includegraphics[width=0.45\textwidth]{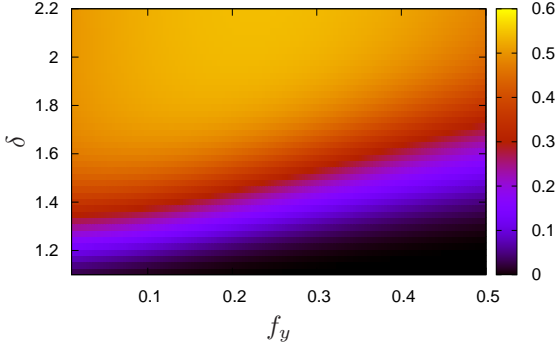}
  \caption{(Color online) EMP $\eta^*$ as a function of the chemical driving force $f_y$ and of the coupling parameter $\delta=\Ubar_1/\Ubar_2$, for the two-dimensional potential (\ref{u2d}), with $T=0.1$, $\lambda_1=0.5$, $\lambda_2=0.1$, $L_1=L_2=3$. $\Pout$ is maximized with respect to $f_x$. }
  \label{fig:2d}
\end{figure}

\begin{figure}[ht]
\psfrag{fy}[ct][ct][1.]{$f_y$}
\psfrag{emp}[ct][ct][1.]{$\eta^*$}
\psfrag{legend1}[cr][cr][0.8]{$\lambda_2=0.8$}
\psfrag{legend2}[cr][cr][0.8]{$\lambda_2=0.1$}
  \centering
  \includegraphics[width=0.45\textwidth]{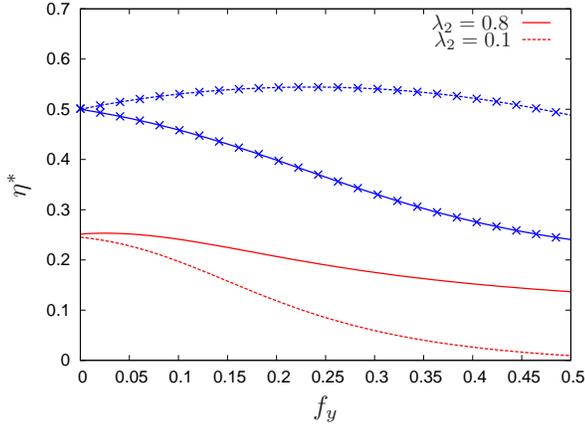}
  \caption{(Color online) EMP $\eta^*$ as a function of $f_y$ for the two-dimensional potential (\ref{u2d}) with $T=0.1$, $\lambda_1=0.5$, $L_1=L_2=3$, $\Ubar_2=1$ and for different values of $\Ubar_1$ and $\lambda_2$. Lines: $\Ubar_1=1.3$, lines and symbols: $\Ubar_1=5$.} 
  \label{fig:2d_diffxb}
\end{figure}

The resulting EMP as a function of $f_y$ and $\delta=\Ubar_1/\Ubar_2$ is shown in fig.~\ref{fig:2d} for $T=0.1$. We note that as the system changes from being a loosely coupled, truly 2-D system ($\delta\simeq1$) to becoming a strongly coupled, effectively 1-D system as $\delta$ is increased, the EMP increases. For sufficiently high $\delta$, the EMP goes beyond the linear response result $1/2$ as expected for the 1-D motion along the $z_2$-direction, since here $\lambda_2<0.5$. Note that the equality $\Pout/\Pin=\wout/\win$ derived for the 1-D case only holds for 2-D systems in the tight-coupling and low-temperature limit where $\vx \propto \vy$. The reason for the above is that we cannot define a single timescale $\mtau$ for motion in a general 2-D potential. Furthermore, we emphasise that in order to obtain $\eta^*>1/2$, an additional structure determined by $U_2$ has to be introduced on top of the pathways defined by a single washboard potential $U_1$.
In figure \ref{fig:2d_diffxb} the EMP is plotted for two different values of $\lambda_2$ and $\Ubar_1$. We see that for large $\Ubar_1$ the EMP goes beyond $1/2$ for $\lambda_2=0.1$, while is stays below $1/2$ for $\lambda_2=0.8$, as expected. For a truly 2-D system, i.e. $\Ubar_2=1.3$ in this case, it is harder to predict the behaviour of the EMP beyond the linear regime as we will illustrate by using a 2-D Kramers model considered in the next paragraph.  

\section{2-D Kramers system}
In the low temperature (high barrier) limit a Kramers description for the 2-D system described by \eqref{u2d} can be developed. In this limit the dynamics can be approximated by a discrete Markov process on a lattice with the lattice points determined by the location of the potential minima. The orientation of the lattice is given by the two directions $a$ and $b$, where the $a$-direction forms an angle $\theta_a$ with the positive x-axis, and the $b$-direction forms and angle $\theta_b$ with the negative $x$-axis. The separation between the points in the two directions is denoted $L_a$ and $L_b$, respectively. In principle, the Kramers formalism allows us to study an arbitrary number of directions. However, the assumption of two directions is the simplest one that allows us to illustrate our arguments, and the results can be compared to the analytical results of the preceding paragraph.

\begin{figure}[h]
\psfrag{fy}[ct][ct][1.]{$f_y$}
\psfrag{emp}[ct][ct][1.]{$\eta^*$}
\psfrag{legend1}[cr][cr][0.8]{$\lambda_2=0.2$}
\psfrag{legend2}[cr][cr][0.8]{$\lambda_2=0.4$}
\psfrag{legend3}[cr][cr][0.8]{$\lambda_2=0.8$}
  \centering
  \includegraphics[width=0.43\textwidth]{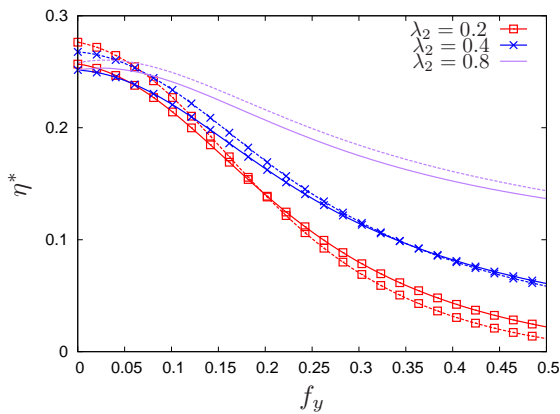}
  \caption{(Color online) EMP $\eta^*$ as a function of $f_y$ for the two-dimensional potential (\ref{u2d}) with $T=0.1$, $\lambda_1=0.5$, $L_1=L_2=3$, $\Ubar_1=1.3$, $\Ubar_2=1$ and for different values of $\lambda_2$. Solid lines: exact, dashed lines: Kramers approximation. Symbols represent different values of $\lambda_2$.}
  \label{fig:kramU11.3}
\end{figure}

\begin{figure}[h]
\psfrag{fy}[ct][ct][1.]{$f_y$}
\psfrag{emp}[ct][ct][1.]{$\eta^*$}
\psfrag{legend1}[cr][cr][0.8]{$\lambda_2=0.2$}
\psfrag{legend2}[cr][cr][0.8]{$\lambda_2=0.4$}
\psfrag{legend3}[cr][cr][0.8]{$\lambda_2=0.8$}
  \centering
  \includegraphics[width=0.43\textwidth]{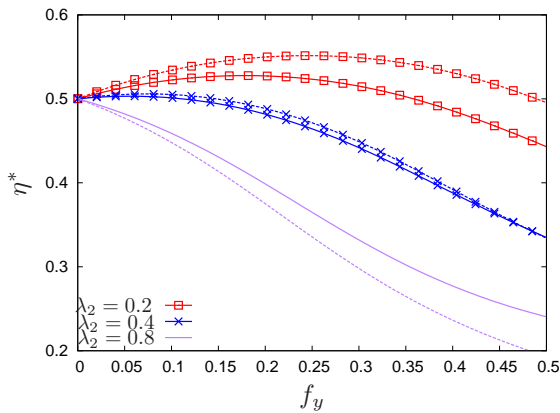}
  \caption{(Color online) EMP $\eta^*$ as a function of $f_y$ for the two-dimensional potential (\ref{u2d}) with $T=0.1$, $\lambda_1=0.5$, $L_1=L_2=3$, $\Ubar_1=5$, $\Ubar_2=1$ and for different values of $\lambda_2$. Solid lines: exact, dashed lines: Kramers approximation. Symbols represent different values of $\lambda_2$.}
  \label{fig:kramU15}
\end{figure}

The four transition rates for the problem become $k^\pm_\alpha=k_0 \exp(-(E_\alpha-f_\alpha^\pm)/T)$, where $\alpha=a,b$, and $k_0$ is a microscopic rate constant assumed to be the same for both directions. The saddle points of the unperturbed potential determine the barrier height $E_\alpha$ and the position $x_\alpha L_\alpha$ of the barrier along the $\alpha$-direction. The quantities $f_\alpha^\pm$ are the changes in the barrier height due to the chemical and mechanical forces, as  given by
\begin{align}
  f_a^+&=(f_x\cos\theta_a + f_y\sin\theta_a)x_aL_a, \\
  f_a^-&=(-f_x\cos\theta_a-f_y\sin\theta_a)(1-x_a)L_a, \\
  f_b^+&=(-f_x\cos\theta_b + f_y\sin\theta_b)x_bL_b, \\
  f_b^-&=(f_x\cos\theta_b-f_y\sin\theta_b)(1-x_b)L_b.
\end{align}
The velocities along the $a$ and $b$ directions become 
  $v_\alpha=(k_\alpha^+-k_\alpha^-)L_\alpha$,
and hence we obtain
\begin{align}
  v_x&=\cos\theta_a v_a - \cos \theta_b v_b \label{eqvxtrig} \\
  v_y&=\sin\theta_a v_a + \sin \theta_b v_b.
\end{align}
In this model we can thus obtain explicit expressions for the EMP optimized with respect to $f_y$. We have studied the expansion $\eta^*=\eta_0+\eta_1f_y+O(f_y^2)$, and for the first term we obtain  
\begin{eqnarray}
 \eta_0 &=&\frac12 \left(L_a^2\cos\theta_a\sin\theta_a-e^\Delta L_b^2\cos\theta_b\sin\theta_b \right)^2\nonumber\\
 &&\times \left\{\left[ (L_a\cos\theta_a)^2+e^\Delta(L_b\cos\theta_b)^2 \right]\right.\nonumber\\&&\quad{}\left.\left[ (L_a\sin\theta_a)^2+e^\Delta(L_b\sin\theta_b)\right] \right.\nonumber\\&&
  \left. \qquad{}+ e^\Delta\left[ L_a L_b\sin(\theta_a +\theta_b)\right]^{2}\right\}^{-1},\label{eq:eta0}
\end{eqnarray}
where $\Delta=(E_a-E_b)/T$. From \eqref{eq:eta0} we obtain $\eta_0\leq 1/2 $, which can be easily shown to hold generally in the linear regime due to the linear structure of the velocities. We also note that $\eta_0$ does not depend on the asymmetry parameters $x_\alpha$. Furthermore, in the limit $\Delta\to\infty$,  eq.~(\ref{eq:eta0}) gives $\eta=1/2$, while from eqs.~(\ref{eqvxtrig}) we obtain that $v_x\propto v_y$. Thus we confirm the result already found numerically for the sawtooth potential, that in the linear regime the EMP for loosely coupled 2-D systems is always smaller than $1/2$ and approaches the limit $1/2$ for the tightly coupled, effectively 1-D system.
The expression for $\eta_1$ is quite long and cumbersome and will therefore not be presented here. In the limit $\Delta\to\infty$ it reduces to
\begin{equation}
  \label{eq:eta1_delta}
  \lim_{\Delta\to\infty} \eta_1= 1/8(1/2-x_b)\sin\theta_bL_b/T,
\end{equation}
as expected from \eqref{eq:seifert_eta_small_fmu}. The sign of $\eta_1$ depends in a complicated way on the parameters of the system: whether the EMP for a 2-D system can rise beyond $1/2$ in the non-linear regime is thus model parameter dependent.    
We have compared the Kramers model with the exact results for a Fourier series expansion for a 2-D sawtooth potential as described in the preceding paragraph. Furthermore, in this study we have included the dependence of the position of the minima and the saddle points on the forces, i.e., $x_\alpha(f_x,f_y)$ and $L_\alpha(f_x,f_y)$. The overall agreement is good as can be observed in figures~\ref{fig:kramU11.3} and~\ref{fig:kramU15}. However, for very asymmetric potentials, e.g., $\lambda_2=0.1$ or $0.9$, the agreement between the exact results and the Kramers approximation is not as good, since our assumption that the microscopic rate $k_0$ is the same along all the directions breaks down. For an effective 1-D system (fig. \ref{fig:kramU15}) the discrepancy is already present for $\lambda_2=0.2$ and $0.8$, since in this case the asymmetry around the minima is more pronounced than for the truly 2-D case. 

\section{Conclusions} 
We have used a Fokker-Planck formalism to study the kinetic and thermodynamic properties of model molecular machines.
In the 1-D case, we exploit the general result that the ratio between the currents along the positive and a negative direction only depends on the imbalance between the mechanical and the ``chemical'' force. We can thus write down a simple expression for the steady-state velocity that only depends on a microscopic timescale for the motion and the splitting probabilities, eq.~(\ref{eq:vst}). Our formalism thus allows us to gain a deeper insight into the connection between the mechanics and the thermodynamics of our model machines. Furthermore, we investigate the specific example of a sawtooth potential. 
We also study a broad class of 2-D potentials in both the continuous and in the Kramers formalism, where we can obtain the exact steady state velocities along the spatial and the chemical directions. We find that for a loosely coupled system, the EMP is always smaller than for a tightly coupled one, where tight coupling corresponds to introducing a ``pathway'' for the particles in the potential, i.e., an effective 1-D system. In the linear regime we obtain that the EMP is always smaller than $1/2$ and approaches $1/2$ as the coupling increases in strength.
\acknowledgments
NG and AI gratefully acknowledge financial support from Lundbeck
Fonden. LP acknowledges the support of FARO and of PRIN 2009PYYZM5.
We thank M. Esposito, D. Lacoste,  and D. Mukamel for interesting discussions.
After the submission of this paper a relevant preprint concerning the present problem has appeared: C. Van den Broeck, N. Kumar,  K. Lindenberg, arXiv:1201.6396


\bibliography{machineRe5}

\end{document}